%% file: paper.tex
\documentclass{sig-alternate}

\input{mymacros}

\makeatletter
\newif\if@restonecol
\makeatother

\usepackage{hhline}
\usepackage[table]{xcolor}
\usepackage{listings}
\usepackage{amsmath, amssymb}
\usepackage{booktabs}
\usepackage{floatrow}
\usepackage[ruled, vlined, linesnumbered]{algorithm2e}
\usepackage{tikz}
\usepackage{fixltx2e}
\usepackage{pgfplots, pgfplotstable}
\usepackage{stmaryrd}

\usetikzlibrary{arrows}
\usetikzlibrary{trees}

\lstdefinestyle{code}{
  language={C},
  basicstyle=\scriptsize\fontfamily{lmvtt},
  numbers=left,                                                                  
  xleftmargin=2.4em,
  frame=single,
  framexleftmargin=2.4em,
  columns=fullflexible,
  numberstyle=\scriptsize
}

\usetikzlibrary{arrows}
\usetikzlibrary{trees}

\tikzstyle{bag} = [text width=9em,
text centered]

\tikzstyle{bag_mod} = [text width=2em,
text centered]

\tikzstyle{bag_rect} = [draw=black,rectangle, black,text width=7em,
text centered]

\tikzstyle{bag1} = [draw=black,rectangle, black,text width=8.5em,
text centered]

\newcommand{\MintHint}{\ensuremath{\mathsf{MintHint}}}
\newcommand{\minthint}{\ensuremath{\mathsf{MintHint}}}

\newcommand{\replace}{\textsf{replace}}

\newcommand{\SF}{\ensuremath{S\!F}}

\newcommand{\SH}{\ensuremath{S\!H}}

\toappear{}

\newcommand{\subparagraph}{}
\usepackage[compact]{titlesec}
\titlespacing{\section}{0pt}{1ex}{1ex}
\titlespacing{\subsection}{0pt}{1ex}{0ex}
\titlespacing{\subsubsection}{0pt}{0.5ex}{0ex}
\begin{document}

\title{{\huge \MintHint}: Automated Synthesis of Repair Hints}

\numberofauthors{1} 
\author{
\begin{tabular}{cc}
\begin{tabular}{c}
\begin{tabular}{ccc}
  Shalini Kaleeswaran & Varun Tulsian & Aditya Kanade\\
\end{tabular}\\ 
\affaddr{Indian Institute of Science}\\
\email{\{shalinik,varuntulsian,kanade\}@csa.iisc.ernet.in}
\end{tabular}
&
\begin{tabular}{c}
  Alessandro Orso\\
  \affaddr{Georgia Institute of Technology}\\
  \email{orso@cc.gatech.edu}
\end{tabular}
\end{tabular}
}

\maketitle

\begin{abstract}
\input{abstract}
\end{abstract}


\input{intro}

\input{overview}

\input{approach}

\input{eval}


\input{relatedwork}

\input{conclusion}

\bibliographystyle{abbrv}
\bibliography{fseref,paper}

\balancecolumns

\end{document}

%% file: mymacros.tex
\usepackage{times}
\usepackage{cite}
\usepackage{graphicx}
\usepackage{epsfig}
\usepackage{xspace}
\usepackage{latexsym}
\usepackage{natbib}
\usepackage[usenames]{color}
\usepackage{multirow}
\usepackage{listings}
\usepackage{textcomp}

\bibpunct{[}{]}{,}{n}{}{}

\def\denseitems{
  \itemsep1pt plus1pt minus1pt
  \parsep0pt plus0pt
  \parskip0pt\topsep0pt}

\makeatletter
\newcommand\tabcaption{\def\@captype{table}\caption}
\makeatother

\newcommand{\ie}{\textit{i.e.,} }
\newcommand{\eg}{\textit{e.g.,} }

\newcommand\opt[1]{}
\newcommand\find[1]{}

\def\denseitems{
  \itemsep1pt plus1pt minus1pt
  \parsep0pt plus0pt
  \parskip0pt
  \topsep0pt
}

\newcommand{\ls}[1]
   {\dimen0=\fontdimen6\the\font 
    \lineskip=#1\dimen0
    \advance\lineskip.5\fontdimen5\the\font
    \advance\lineskip-\dimen0
    \lineskiplimit=.9\lineskip
    \baselineskip=\lineskip
    \advance\baselineskip\dimen0
    \normallineskip\lineskip
    \normallineskiplimit\lineskiplimit
    \normalbaselineskip\baselineskip
    \ignorespaces
   }

\newenvironment{smalldescription}{
   \setlength{\topsep}{0pt}
   \setlength{\partopsep}{0pt}
   \setlength{\parskip}{0pt}
   \begin{description}
   \setlength{\leftmargin}{.2in}
   \setlength{\parsep}{0pt}
   \setlength{\parskip}{0pt}
   \setlength{\itemsep}{0pt}}{\end{description}}

\newcommand{\myskip}[1]{}
\newcommand{\lineone}{$80$}
\newcommand{\linetwo}{$77$}
\newcommand{\linethree}{$70$}
\newcommand{\linefour}{$73$}
\newcommand{\linefive}{$502$}



%% file: abstract.tex
Being able to automatically repair programs is at the same time a very
compelling vision and an extremely challenging task.  In this paper,
we present \MintHint, a novel technique for program repair that is a
departure from most of today's approaches. Instead of trying to fully
automate program repair, which is often an unachievable goal,
\MintHint\ performs statistical correlation analysis to identify
expressions that are likely to occur in the repaired code and
generates, using pattern-matching based synthesis, \textit{repair
  hints} from these expressions. Intuitively, these hints suggest how
to rectify a faulty statement and help developers find a complete,
actual repair. \MintHint\ can address a variety of common faults,
including incorrect, spurious, and even missing expressions.

We also present an empirical evaluation of \MintHint\ that consists of
two main parts. The first part is a user study that shows that, when
debugging, developers' productivity can improve manyfold with the use
of repair hints---compared to having only traditional fault
localization information.  The second part consists of applying
\MintHint\ to several faults of a widely used Unix utility program to
further assess the effectiveness of the approach.  Our
results show that \MintHint\ performs well even in situations,
seen frequently in practice, where (1) the repair space searched does not contain the exact
repair, and (2) the operational specification obtained from the
test cases for repair is incomplete or even imprecise---which
can be challenging for approaches aiming at fully automated repair.


%% file: intro.tex
\section{Introduction}
\label{sec:introduction}

Debugging is an expensive activity that can be responsible for a
significant part of the cost of software maintenance~\cite{vessey85}.
It is therefore not surprising that researchers and practitioners
alike have invested a great deal of effort in developing techniques
that can improve the efficiency and effectiveness of debugging (\eg
\cite{groce04, ball03, jones2002, renieris03, zhang06icse,
  DBLP:conf/icse/CleveZ05, samimi2012automated, chandra2011,
  nguyen2013, legoues-tse2012, pei2011code, demsky2006inference}).  In
particular, in recent years, there has been a growing interest in
automated program repair techniques (\eg \cite{legoues-tse2012,
  pei2011code, nguyen2013, samimi2012automated, demsky2006inference,
  chandra2011}). Although these techniques have been shown to be
effective, they suffer from one or more of the following limitations.
First, some techniques rely on the existence of a specification for
the program being debugged, which is rarely the case in practice.
Second, techniques that do not rely on specifications tend to
``overfit'' the repair to the set of existing test cases, which is
likely to affect the general validity of the repair.  Third, because
they are looking for a complete repair, most existing techniques must
perform a search over a repair space that is large enough to include
the (unknown) repair. For non-trivial repairs, this can make the
technique either ineffective (if the bound on the repair space used by
the tool is too small) or too expensive to be used in practice (if the
bound used/required is too large).

\begin{figure}[!t]
\begin{lstlisting}[language=C, firstnumber=65]
char esc(char *s, int *i) {
   char result; 
   if (s[*i] != ESCAPE)
      result = s[*i];
   else 
      if (s[*i] == ENDSTR) // Localization rank 3
         result = ESCAPE;
      else {
	*i = *i + 1; // Localization rank 4
        if (s[*i] == 'n')
           result = NEWLINE;
        else
           if (s[*i] == 't') // Localization rank 2
              result = TAB;
           else 
              result = s[*i]; // Localization rank 1
        }
   return result;
}
\end{lstlisting}
\vspace{-8pt}
\caption{Function \textsf{esc} from program \replace\ (version $23$),
  which contains a faulty statement at line \linethree. The ones shown are
  the line numbers of function \textsf{esc} in the actual code.}
\vspace{-16pt}
\label{fig:example}
\end{figure}

To address these limitations of existing techniques, in this paper we
propose \MintHint, a novel, semi-automated approach to program repair.
\MintHint\ is a departure from most of today's program repair
techniques as it does \emph{not} try to find a complete repair,
which we have observed to be an unachievable goal in many, if not
most, cases due to technical and practical reasons. Instead,
\MintHint\ aims to \textit{generate repair hints that suggest how to
  rectify a faulty statement and help developers find a complete,
  actual repair}.

As an example, consider function \textsf{esc} from a faulty version of
program \replace~\cite{siemens}, shown in Figure~\ref{fig:example}.
Function \textsf{esc} takes as input a
string and an index into the string and checks whether the character
at the index is a special character (such as a newline or a tab). If
so, it returns a program-specific constant that represents the special
character.  The fault is at line~\linethree, where the branch
predicate should be \textsf{s[*i+1] == ENDSTR}, but an incorrect array
index, \textsf{*i}, is used instead.  Given this faulty program
and a set of test cases for the program that trigger the fault (\ie at
least one test in the set fails due to this fault), \MintHint\ would
produce the following hint:

\begin{center}
  \textbf{Replace} \textsf{s[*i] == ENDSTR} by \textsf{s[*i+1] == ENDSTR}
\end{center}

\noindent 
Developers would use this hint as guidance while modifying the
original code to arrive at a repair. In this specific example,
developers would simply make the suggested change and obtain the
repaired version.  In the more general case, as we will show in
Section~\ref{sec:eval}, hints are not necessarily complete repairs, but rather
suggestions on how to generate such repairs.

In the common case in which more than one statement is suspected to be
faulty, \MintHint\ would \emph{algorithmically} generate repair hints
for all of them.  It would then rank the generated hints to help
developers prioritize their efforts.

To generate hints, \MintHint\ operates in four steps. The
\textit{first step} identifies potentially faulty statements by
leveraging an existing fault localization technique that requires only
a test suite (\eg \cite{jones2002, ochiai, zoltar}).  The subsequent
steps are performed for each of the identified statements. The
\textit{second step} derives a state transformer, that is, a function
that (1) is defined for all program states that reach the faulty
statement in the given test suite and (2) produces the right output
state for each of them. This step leverages dynamic symbolic execution
techniques (\eg~\cite{DBLP:conf/osdi/CadarDE08}).  The \textit{third
  step} explores a repair space and tries to identify and rank,
through a statistical correlation analysis~\cite{bookCramer,bookRank},
expressions in the space that are likely to occur in the repaired
statement, using the state transformer derived in the second step. (To
the best of our knowledge, this paper is the first one to apply this
form of statistical reasoning to programs.)  Finally, the
\textit{fourth step} of the approach synthesizes repair hints by
pattern matching~\cite{DBLP:journals/ipl/ZhangSS92} 
the expressions computed in the previous step with
those in the faulty statement. Together these steps
result in a sophisticated technique that suitably \emph{combines symbolic,
  statistical, and syntactic reasoning} to help in program repair.

\MintHint\ synthesizes five types of hints that suggest (1) insertion,
(2) replacement, (3) removal, or (4) retention of expressions, and (5)
combinations of these. As shown in Table~\ref{table:hints-summary},
these hints are applicable to many types of common faults, ranging
from incorrect or spurious expressions to missing expressions, to
combinations of these. Moreover, \MintHint\ can handle faults in a
variety of program constructs, such as assignments, conditionals,
switch statements, loop headers, return statements, and statements
with ternary expressions. The main restriction, for the technique
presented in this paper, is that the fault has to involve a single
statement and, in the case of an assignment, it must be on the
right-hand side (RHS) expression.  However, we believe that \MintHint\ can
be extended to address other situations, \eg~where the
left-hand side variable in an assignment is faulty or the fault
spans multiple statements.

\MintHint\ overcomes the main limitations of existing techniques that
we listed earlier in the following ways. First, it does not rely on a
formal specification; it instead derives an operational specification
(\ie a state transformer) from the test cases available. Second,
approaches that aim at deriving complete repair, typically use
equality with the state transformer (or an analogous entity) as a
criterion for selecting a candidate repair (\eg~\cite{legoues-tse2012,
  nguyen2013}).  
The statistical correlation used in \MintHint\ is a
\emph{more relaxed and robust notion than equality} and can thus be more
effective in identifying which expressions are \emph{likely} to be
part of the repaired code; this allows \MintHint\ to generate more
general repairs and to be effective in the presence of incomplete or
even imperfect (\ie noisy) data.  Third, since \MintHint\ looks 
for building blocks of repair (rather than the complete repair itself)
and then combines them algorithmically to generate compound hints, it can
generate useful \textit{actionable} hints even when exploring an
incomplete repair space.

To evaluate \MintHint, we developed a prototype tool that implements
our approach for C programs and performed a \emph{user study} using
programs from the Siemens benchmark~\cite{siemens}. Specifically, 
the study consisted of two phases: control and experimental.
In both phases, a user was provided a single repair task along with
fault localization information and a test suite.
In the experimental phase, in addition, repair hints generated by \MintHint\
were supplied. The tasks given to a user in the two phases were independent.
In all $10$ users were involved in the study. 
Without repair hints, 
only $6$ of them completed their task within $2$h. With repair
hints, \emph{all} users could complete their task within the same
time limit. For example, the fault described in Figure~\ref{fig:example}
was part of the user study but without hints the user could
not repair it within $2$h, whereas with hints another user repaired it easily.
Moreover, for the tasks completed in both the phases,
in the experimental phase, with hints, they were completed
over $5$ times faster. 

In addition to the user
study, we also performed a \emph{case study} using a commonly used Unix
utility, namely, the stream editor \textsf{sed}. We considered $6$ faulty
versions of \textsf{sed}. On one of them symbolic execution timed out
and hence it was not subjected to hint generation.
On $3$ of the remaining $5$ faulty versions, \MintHint\ provided
hints that immediately lead to repair. In one additional case,
the hints when applied to the program, resulted in a partial repair
in which all passing tests continued to pass but several failing
tests started passing. Further, \MintHint\ was
able to synthesize useful hints even in the many cases across
the user study and the case study in which (1) the
repair space considered did \textit{not} contain the repaired version
of the faulty expression, or (2) the state transformer contained
imperfect data. 

In summary, the main contributions of this paper are:

\vspace{-8pt}
\begin{itemize}\denseitems

\item The definition of a novel technique for program repair, called
  \MintHint, that combines symbolic, statistical, and syntactic
  reasoning and overcomes some of the main limitations of existing
  techniques by focusing on generating repair hints, rather than
  complete repairs.

\item 
  An implementation of \MintHint\ that
  can perform automated synthesis of repair hints for C programs.

\item A user study that evaluates the effect of repair hints on
  developers' productivity. This is one of the few user studies
  performed in the area of program repair and debugging in general.

\item A further evaluation of \MintHint's effectiveness on a case
  study involving several faulty versions of a 
   Unix utility.

\end{itemize}

\begin{table}
\centering
\begin{tabular}{|l|l|}
\hline
\emph{Nature of hint} & \emph{Targeted fault}\\
\hline
Insert & Missing expressions\\
\hline
Replace & Incorrect operator, constant, variable, etc.\\
\hline
Remove & Spurious expressions\\
\hline
Retain & Eliminating false positives\\
\hline
Compound & One or more occurrences of above\\
\hline
\end{tabular}
\vspace{-8pt}
\caption{Nature of hints and targeted faults.}
\vspace{-8pt}
\label{table:hints-summary}
\end{table}



%% file: overview.tex
\section{Overview}
\label{sec:overview}

\begin{figure*}[t]
  \centering
  \includegraphics[width=.86\textwidth]{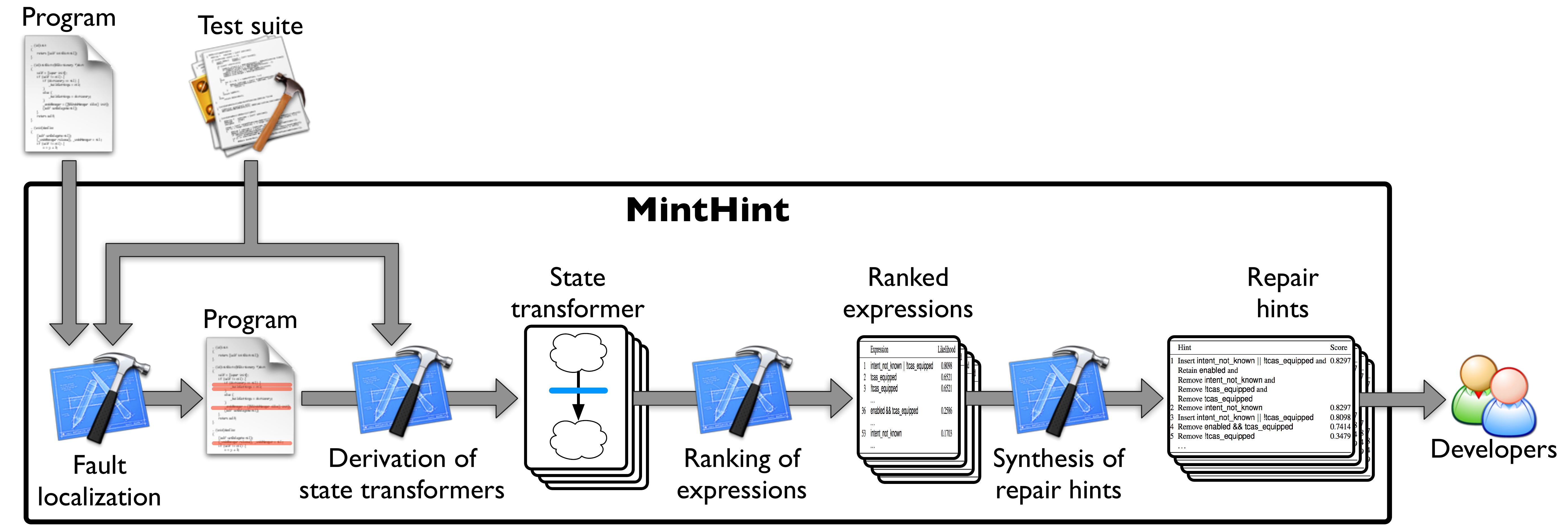}
  \caption{High-level view of the \MintHint\ approach.}
  \vspace*{-16pt}
  \label{fig:overview}
\end{figure*}

Figure~\ref{fig:overview} provides a high-level view of \MintHint.  As
the figure shows, \MintHint\ takes as input a faulty program and a
test suite---where at least one test case triggers the fault in the
program, and thus fails---and produces a list of repair hints in four
steps. We now describe these four steps using the example we discussed
in the Introduction, shown in Figure~\ref{fig:example}.

\paragraph*{Step 1: Fault localization.}

This is a preliminary step, whose goal is to provide the hint
generation algorithm with a list of possibly faulty statements to be
repaired. To compute this list, \MintHint\ can leverage any existing
fault localization approach. In our current implementation, we use the
Ochiai approach as implemented in the Zoltar tool~\cite{zoltar}, which
uses the test suite and performs spectra-based fault localization. For
our example, the localization tool outputs a ranked list of suspicious
statements: \lineone, \linetwo, \linethree, \linefour, \linefive, and
so on. 
In its subsequent steps, \MintHint\ runs the hint generation algorithm
on each statement in the fault localization list \emph{independently},
up to a given threshold. In the following discussion, we use
line~\linethree\ to illustrate the remaining steps of \MintHint.

\begin{table}
{\small
\begin{tabular}{@{}cccccc@{}}
\toprule
\multicolumn{2}{c}{\emph{Input state}} &
\emph{Output state} & \multicolumn{3}{c}{\emph{Values of expressions over input state}}\\
\cmidrule{1-2}\cmidrule(lr){3-3}\cmidrule(lr){4-6}
\textsf{s} & \textsf{*i}  & \textsf{branch$_0$} &
\textsf{s[*i] ==} & \textsf{s[*i] ==} & \textsf{s[*i+1] ==}\\
& & & \textsf{s[*i+1]} & \textsf{ENDSTR} & \textsf{ENDSTR} \\
\midrule
"@\%\&a" & $0$ & \emph{false} & false & false & \emph{false}\\
"@n@" & $0$ & \emph{false} & false & false & \emph{false}\\
"@n@" & $2$ & \emph{true} & false & false & \emph{true}\\
"\%@@" & $1$ & \emph{false} & true & false & \emph{false}\\
"V@" & $1$ & \emph{true} & false & false & \emph{true}\\
\bottomrule
\end{tabular}
}
\vspace{-8pt}
\caption{An example state transformer and values of some 
expressions over the input states.}
\label{table:st-trans}
\vspace{-8pt}
\end{table}

\paragraph*{Step 2: Derivation of state transformers.}

The next step towards repair is to infer the specification of what
would be the correct statement at line~\linethree. To do so, in the
absence of an actual specification, \MintHint\ uses the test suite
provided for the program to infer an operational specification for the
statement in the form of a state transformer. Informally, this
\textit{state transformer} is a function that, given an input state
(\ie valuation to variables) at the (potentially) faulty statement,
produces an output state that would make every test in the test suite
pass, including the failing ones. For passing tests, state
transformers can be easily computed by simply observing input and
output states during the execution of the tests. For failing tests, as
we discuss in detail in Section~\ref{sec:tests-likelihood}, \MintHint\
computes state transformers using dynamic symbolic execution and
constraint solving~\cite{DBLP:conf/osdi/CadarDE08}. Let us illustrate
this step using the statement at line~\linethree. Because the
statement is a conditional statement, \MintHint\ uses a fresh boolean
variable, say \textsf{branch$_0$}, to represent the outcome of the
branch predicate.  That is, the statement is
transformed into an assignment \textsf{branch$_0$ = s[*i] == ENDSTR},
in which the fault is on the RHS, and the
branch predicate becomes the variable \textsf{branch$_0$}. 
Table~\ref{table:st-trans} shows some entries in 
the state transformer of the statement: (1)~the string array \textsf{s}
and index expression \textsf{*i} in the input state and
(2)~the LHS variable \textsf{branch$_0$} in the output state.
The values of all variables except \textsf{branch$_0$} remain unchanged
between a pair of input/output states, and hence are not shown again.
The value of \textsf{branch$_0$} in the first row is obtained from a passing test by concrete execution
and the remaining are obtained from failing tests by symbolic execution.
These are the values that \textsf{branch$_0$} is \emph{expected} to take
in the repaired version, on the respective input states.

\paragraph*{Step 3: Ranking of expressions.}

The goal of this step is to identify \emph{syntactic building blocks} (\ie
expressions) for constructing an RHS expression compliant with the
computed state transformer. To do so, \MintHint\ searches the solution
space, called the \emph{repair space}, for expressions whose values
over the input state of the state transformer are statistically
correlated with the corresponding output values produced by the state
transformer.  More precisely, \MintHint\ interprets \emph{correlation
  coefficients}, which represent the numerical measure of the strength
of a statistical correlation~\cite{bookCramer, bookRank}, as the
\emph{likelihood} of the expression to occur in the repaired RHS.
For our example, \MintHint\ would compute correlations between the
values of the candidate expressions and the expected value of
\textsf{branch$_0$} at line \linethree. 
For some expressions from the repair space and the faulty RHS itself,
Table~\ref{table:st-trans} gives the values over the input states.
\textsf{ENDSTR} is the null character indicating end-of-string.
As can be seen, the values of \textsf{s[*i+1] == ENDSTR} are 
highly correlated with the expected values of \textsf{branch$_0$}
but the same is not true for the other expressions.
Table~\ref{table:corr} shows some expressions in the
repair space ranked by decreasing value of likelihood.

\begin{table}[t]
{\small
\begin{tabular}{r l c}
\toprule
\emph{Rank} & \emph{Expression} & \emph{Likelihood} \\
\midrule
  1 & \textsf{s[*i+1] == ENDSTR} & 0.62\\
  2 & \textsf{s[*i+1] <= ENDSTR} & 0.62\\
    & \ldots &  \\
  7 & \textsf{s[*i+1] != ENDSTR} & 0.62\\
    & \ldots & \\
 23 & \textsf{s[1] == ENDSTR} & 0.43\\
    & \ldots & \\
488 & \textsf{s[*i] == ENDSTR} & 0\\
& \ldots & \\
 \bottomrule
\end{tabular}
}
\vspace{-8pt}
\caption{A partial list of expressions ranked by the likelihood of
  occurrence in the repaired RHS.}
\vspace{-8pt}
\label{table:corr}
\end{table}

\paragraph*{Step 4: Synthesis of repair hints.}

After producing a ranked list of expressions, \MintHint\ analyzes this list
to synthesize an \textit{actionable list} of repair hints. \MintHint\
generates two types of hints: simple and compound. A \emph{simple
  hint} is a single program transformation, whereas a \emph{compound
  hint} is a set of program transformations.  For our example, the
 hints synthesized by \MintHint\ for some of
the statements identified by the fault localization tool are given in
Table~\ref{table:exhints}. Note that \MintHint\ may generate more
than one hint for the same statement (\eg~statement $70$ in Table~\ref{table:exhints}).
We now explain how \MintHint\ generates such hints.

\begin{table}[t]
{\small
\begin{tabular}{ccll}
\toprule
\emph{Rank} & \emph{Statement} & \emph{Hint} & \emph{Score}\\
\midrule
1 & \linethree & \textbf{Replace} \textsf{s[*i] == ENDSTR} & $1$\\
  &      & with \textsf{s[*i+1] == ENDSTR} & \\
2 & \linethree & \textbf{Replace} \textsf{s[*i] == ENDSTR} & $1$\\
  &      & with \textsf{s[1] == ENDSTR} & \\
3 & \linethree & \textbf{Insert} \textsf{s[*i+1] <= ENDSTR} & $1$\\
  &      & and \textbf{Remove} \textsf{s[*i] == ENDSTR} & \\
4 & \linetwo & \textbf{Retain} the statement & $1$\\
5 & \lineone & \textbf{Retain} the statement & $1$\\
6 & \linefive & \textbf{Retain} the statement & $1$\\
7 & \linethree & \textbf{Insert} \textsf{s[*i+1] != ENDSTR} & $0.62$ \\
\bottomrule
\end{tabular}
}
\vspace{-8pt}
\caption{Repair hints synthesized by \MintHint.}
\vspace{-8pt}
\label{table:exhints}%
\end{table}

\vspace{2mm}
\noindent
For \textbf{simple hints}, \MintHint\ iteratively selects expressions from
the repair space of each statement individually such that
(1)~their likelihood values are above a threshold and 
(2)~the statistical correlation among themselves is low.
That is, they form a set of expressions such that \emph{any one}
expression in the set is likely to appear in the repaired RHS.
In our example, \MintHint\ would select \textsf{s[*i+1]
  != ENDSTR} and \textsf{s[1] == ENDSTR}.  We discuss the algorithm
for selection of expressions in detail in Section~\ref{sec:synth}.

After selecting these expressions, \MintHint\ \emph{pattern
  matches}~\cite{DBLP:journals/ipl/ZhangSS92} each of the selected
expressions with the faulty RHS.  Based on the edit distance between
an expression and the subexpressions in the faulty RHS, \MintHint\
determines the nature of the hint to be generated for the expression. 
If the edit distance is less than or equal
to a threshold chosen heuristically ($2$ in our current formulation),
\MintHint\ suggests the replacement of the matching expression in the
faulty RHS.  Otherwise, if the edit distance is greater than the
threshold, it suggests an insertion. Expression \textsf{s[1] ==
  ENDSTR}, for example, is at edit distance $2$ from the faulty RHS's
expression \textsf{s[*i] == ENDSTR}, so \MintHint\ synthesizes a
replace hint for the RHS (position $2$ in Table~\ref{table:exhints}). The
edit distance for \textsf{s[*i+1] != ENDSTR}, conversely, is greater
than $2$, and \MintHint\ synthesizes an insert hint for this expression
(position $7$ in Table~\ref{table:exhints}).

If the edit distance of an expression selected through likelihood value
is zero, that is, the expression
already occurs as a subexpression in the faulty RHS, \MintHint\ deems
the RHS's subexpression as \emph{unlikely to be faulty} and generates a retain
hint for it. In our example, this happens for lines
\lineone, \linetwo, and \linefive\ (positions $5$, $4$, and $6$, respectively). 
In all three cases, \MintHint\
finds that the existing RHS is the most likely expression to
appear in the repaired version and hence synthesizes the retain hints 
for the statements. In fact, there is no way to repair
the fault through simple modifications of these statements. We can say
that retain hints can help developers localize a fault by eliminating
some spurious statements (\ie false positives) returned by the fault
localization tool.
%

\MintHint\ ranks hints based on their \emph{score}.  Intuitively, the
score indicates the confidence in the applicability of the hint and is
derived from the likelihood values of the expressions involved. In the
case of a replace hint, the score is the maximum of the likelihood of
the expression being used for replacement and one minus the likelihood
of the expression being replaced. For the replace hint at position $2$
in Table~\ref{table:exhints}, for instance, the score is $\max(0.43,
1-0)$, where $0$ is the likelihood value of the faulty RHS which is
being replaced (see Table~\ref{table:corr}).  For insert and retain
hints, the score is the likelihood of the expression being
inserted or retained (\eg~$0.62$ for the insert hint at
position $7$ in Table~\ref{table:exhints}).  For a remove hint,
it is one minus the likelihood of the expression being removed.

\vspace{2mm}
\noindent Whereas simple hints can address faults that can be repaired
through a single syntactic transformation, \textbf{compound hints} can
help repair more complex faults---faults that require more than one
program transformation to be repaired or more refined pattern matching. 
Further, if the repair space
contains only building blocks of the repaired RHS, but not the
repaired RHS itself, then compound hints---obtained by algorithmically combining the building 
blocks---bring the repair hints closer to the actual repair.

\MintHint\ synthesizes compound hints by first computing 
sets of expressions such that (1)~within each set, the likelihood values 
of the expressions are above a threshold and 
(2)~all expressions in the repair space which are
likely to appear in the repaired RHS \emph{together} are included in the same set.
This computation uses a variant of correlation 
coefficients, called \emph{partial correlation coefficients},
and a more refined pattern matching.
We differ the detailed discussion on this to Sections~\ref{sec:tests-likelihood} and~\ref{sec:synth}.
  In our example, \MintHint\ would compute two such sets:
\{\textsf{s[*i+1] == ENDSTR}\} and \{\textsf{s[*i+1] <= ENDSTR}\}.
The selection criterion used here successfully identifies the 
required expression \textsf{s[*i+1] == ENDSTR}.
Each of these sets is singleton in this case.
In general, however, the sets would contain more than one expression
(\eg~see task 10 in Section~\ref{sec:user-study}).

After computing a set of expressions that may occur together in a
repair, \MintHint\ synthesizes actual compound hint 
using the edit distance of each expression
in the set with the faulty RHS. 
For the first set in
our example, \{\textsf{s[*i+1] == ENDSTR}\}, \MintHint\ generates the
replace hint at position $1$ in Table~\ref{table:exhints}. For the
second set, \{\textsf{s[*i+1] <= ENDSTR}\}, it generates the hint at
position $3$ in Table~\ref{table:exhints}, which has two \emph{constituent hints}:
an insert hint for the expression in the set together with a remove hint. \MintHint\ adds
the remove hint because, in this compound hint, there is no constituent
hint which suggests retention or replacement of the faulty RHS. 

To compute the score for a compound hint, \MintHint\ computes the
maximum of the scores of the constituent hints. After computing all
hints and their scores, \MintHint\ reports the
hints to the developer, ordered by score (as in Table~\ref{table:exhints}). 

\vspace{2mm}
\noindent
After getting the hints produced by \MintHint, developers can manually
{\bf apply the hints} by modifying the potential faulty statement according
to the hints. For example, in the case of a replace hint, developers
will have to replace a subexpression with a suggested one. For a
remove hint, developer will have to remove the subexpression in the
hint. They will also have to suitably remove the operator(s) around
the subexpression, or guess another expression to fill the hole, so as
to obtain a well-formed resulting expression. Similarly, for an insert
hint, developers will have to combine the subexpression in the hint
with the existing expression and select an appropriate operator and
place for insertion. 


%% file: approach.tex
\begin{algorithm}[t]
{\small
 \SetAlgoNoLine
 \DontPrintSemicolon
 \SetInd{0.5em}{0.5em}
 \KwIn{Program $P$, test suite $T$, the number of faulty stmts $k$,
the bound on the size of exps in the repair space $m$}
 \KwOut{Ranked list of repair hints for the faulty statements}
 \label{topAlgo}
 
 \Begin{
     $\SF \leftarrow \mathsf{localize\_faults}(P,T,k)$ \tcp{\small Localize faults}
     \label{line:localize}
     $Hints \leftarrow \emptyset$ \tcp{\small Initialize the set of hints}
   \ForEach{$F \in \SF$}{
     \label{loop-start}
   \tcp{\small Derive the state transformer}
   Let $F$ be of the form $x := e$\;
   $f \leftarrow \mathsf{st\_trans}(P, F, T)$\;
     \label{line:state-transformer}
   Let $f$ be represented as an array $[(\sigma_1,\sigma'_1),\ldots,(\sigma_n,\sigma'_n)]$\;
   \label{f-as-array}
   \tcp{\small Enumerate expressions in repair space}
   $S \leftarrow \mathsf{subexps}(e)$\;
   \label{compute-s}
   $V \leftarrow \mathsf{vars\_in\_scope}(P,F)$;
   $E \leftarrow \mathsf{enum\_exps}(G, V, m)$\;
   \label{compute-e}
   \tcp{\small Generate the data}
   $D(x) \leftarrow [\sigma'_1(x),\ldots,\sigma'_n(x)]$\;
   \label{dataset1}
   \lForEach{$e' \in S \cup E$}{
     \label{dataset2}
     $D(e') \leftarrow [\sigma_1(e'), \ldots, \sigma_n(e')]$
   }
   \tcp{\small Synthesize hints}
   $Hints \leftarrow Hints \cup \mathsf{MintSimpleHints}(F, D, S, E)$\;
   $Hints \leftarrow Hints \cup \mathsf{MintCompoundHints}(F, D, S, E)$\;
   \label{compound-hints}
 }
   \Return{$\mathsf{sort}(Hints)$}
   \label{rank}
 }
}
 \caption{Algorithm $\mathsf{MintHint}$}
 \label{algo:minthint}
\end{algorithm}

\section{Algorithm}
\label{sec:approach}

The \MintHint\ algorithm is presented in
Algorithm~\ref{algo:minthint}. As we discussed in the previous
section, the input to the algorithm is a program $P$ and a test suite
$T$ such that the program fails on at least one of the tests in $T$.
\MintHint\ also takes a threshold $k$ on the number of faulty
statements to be considered for repair and a bound $m$ on the size of
the expressions in the repair space.

The first step of \MintHint\ (line~\ref{line:localize}) is to use the
test suite to perform fault localization, as indicated by the function
$\mathsf{localize\_faults}$.  All statements in the fault localization
list $SF$ are then processed in a loop (starting at
line~\ref{loop-start}). Function $\mathsf{st\_trans}$ at
line~\ref{line:state-transformer} computes the state transformer for a
faulty statement $F$. Sets $S$ and $E$, computed at
lines~\ref{compute-s} and~\ref{compute-e}, contain the subexpressions
of the faulty RHS and the extra expressions that should be considered
when searching for repairs. The function $D$ maps expressions
from the repair space and the LHS variable to their values 
according to the state transformer (similar to the column values in
Table~\ref{table:st-trans}).  Two separate algorithms,
$\mathsf{MintSimpleHints}$ and $\mathsf{MintCompoundHints}$ are used
for synthesizing simple and compound hints. Finally, the hints across
all statements are sorted by their scores (line~\ref{rank}).

In the following discussion, for simplicity, we consider statement $F$
to be of the form $x := e$. 
A conditional statement can be rewritten in this form similar to 
the transformation of line~\linethree\ in Section~\ref{sec:overview}.
\MintHint\ can 
handle cases in which the potentially faulty statement is a loop
header of the form \textsf{for(init, cond, upd)}, where \textsf{init}
initializes the loop counter(s), \textsf{cond} is the loop termination
condition, and \textsf{upd} is the update of the loop counter(s).  In
these cases, the fault could be in any of these three components, so
\MintHint\ spawns three different tasks---one each for \textsf{init},
\textsf{cond}, and \textsf{upd}. \MintHint\ treats an
assignment with a ternary RHS expression as a conditional statement in
which the fault is in the branch predicate or in one of the
assignments in the branches. \MintHint\ can handle other constructs,
such as \textsf{switch} or \textsf{return}, in analogous ways.  

We now present the different steps of the algorithm (summarized above)
in detail by discussing how \MintHint\ goes (1) from the program's
test suite to expressions that may help repair the fault and (2) from
these expressions to actual repair hints.

\subsection{From Tests to Likelihood of Expressions}
\label{sec:tests-likelihood}

This first part of the \MintHint\ algorithm is based on \emph{symbolic and
  statistical analysis} and is performed in several phases. We discuss
each of these phases separately.

\subsubsection{Fault Localization}
\label{sec:localization}

Function $\mathsf{localize\_faults}$ leverages an existing fault
localization technique to compute a set $\SF$ of potentially faulty
statements.  Any fault localization technique that requires only the
program $P$ and a test suite $T$ can be used here
(\eg~\cite{jones2002, ochiai, zoltar}).  These techniques, called
\emph{spectra-based} techniques, gather runtime data on the execution
of passing and failing tests and use data clustering techniques to
rank statements by their likelihood of being faulty.

Given this ranked list, \MintHint\ analyzes the top $k$ statements in
the list, where $k$ is a threshold set by the user that can be
increased progressively, if necessary.  We recall that, analogously to
other existing repair approaches (\eg~\cite{nguyen2013,
  legoues-tse2012}), \MintHint\ currently assumes that faults can be
repaired by changing a single statement.  Therefore, \MintHint\
synthesizes repair hints for each statement $F$ in $\SF$, the set of
potentially faulty statements, independently.

\subsubsection{Derivation of State Transformers}
\label{sec:state-trans}

For each potentially faulty statement $F$, function
$\mathsf{st\_trans}$ derives a \emph{state transformer} $f$ that, when
substituted to $F$, makes the program produce the correct output for
each test in $T$. $f$ is a function from program states to program
states, where a \emph{program state} $\sigma$ is a mapping from the
variables in scope at the faulty statement to appropriately typed
values.  More formally, in Algorithm~\ref{algo:minthint} $f$ is an
array of pairs of input/output states: $f =
[(\sigma_1,\sigma'_1),\ldots,(\sigma_n,\sigma'_n)]$
(line~\ref{f-as-array}).  Notationally, an unprimed state is an input
state (at $F$), and a primed state is the corresponding output state.

Function $f$ is defined for states that can be witnessed at $F$, given
the inputs in $T$. Tests in $T$ that do not execute $F$ are ignored
when computing state transformers.  For each passing test traversing
$F$, \MintHint\ runs the program and collects the input/output states
at $F$.  Conversely, for failing tests that traverse $F$, \MintHint\
(1) makes the LHS variable $x$ symbolic, (2) uses a \emph{symbolic
  execution algorithm with constraint solving} to obtain a value of
$x$ that makes the program produce the correct output, and (3) reruns
the program concretely using the so computed value for $x$ (instead of
the original value of the RHS $e$). The input/output states at $F$ in
this concrete execution give the mapping $f$ for the failing test.  
\MintHint\ also handles the cases where the faulty statement
is executed more than once in a passing or failing test.

Symbolic execution may fail to obtain a value for $x$ that results in
the correct output.  If all failing tests of $F$ still fail during
symbolic execution (\ie the program fails irrespective of $F$), it is
possible that $F$ may not be the faulty statement. 
In such cases, \MintHint\ generates a ``retain the statement'' hint.
\MintHint\ also sets timeout for symbolic execution
and if the symbolic execution times out on every failing test
within the timeout threshold then \MintHint\ discards the statement
and does not generate any hints.


In its subsequent phases, \MintHint\ treats the computed state
transformer $f$ for each potentially faulty statement $F$ as an
\emph{operational specification} for the repair, thus eliminating the
need to have a formal specification.

\subsubsection{Ranking of Expressions}
\label{sec:ranking}

Using the state transformer $f$ as a specification for statement $F$,
\MintHint\ ranks expressions in the repair space of $F$ according to
their likelihood of occurring on the RHS of the repaired version of
$F$.  The repair space to be searched over can be obtained in several ways, such as by
enumerating expressions over variables in scope or mining expressions
that occur elsewhere in the program (similar to what is done
in~\cite{legoues-tse2012}).  Presently, \MintHint\ uses the former
approach. More precisely, function $\mathsf{vars\_in\_scope}$, in
Algorithm~\ref{algo:minthint}, computes the set $V$ of variables in
scope at $F$. Then, function $\mathsf{enum\_exps}$ enumerates the
expressions of size up to $m$ (a user-defined threshold) over $V$
according to the grammar of expressions in the programming language in
which $P$ is implemented. Let $E$ be the set of these expressions.

\MintHint\ also includes in the repair space the set of subexpressions
of $e$, the RHS expression, including $e$ itself. It does so because
it evaluates whether they are likely to be faulty or not
\emph{independent} of the fault localization results.  This set, which
we call $S$, is computed by function $\mathsf{subexps}$
(line~\ref{compute-s}).  The repair space is thus defined as $E \cup
S$.

We recall that a program state $\sigma$ is a mapping from variables to
values.  This mapping can be extended naturally to expressions.  In
particular, if $e \equiv e' \; op\; e''$ then $\sigma(e) = \llbracket
op \rrbracket(\sigma(e'),\sigma(e''))$.  
Let $D(x) = [\sigma'_1(x),\ldots,\sigma'_n(x)]$ be the
values of the LHS variable $x$ over the output states defined by state
transformer $f$ (line~\ref{dataset1}).  Similarly, for an expression
$e'$ in the repair space, let $D(e') =
[\sigma_1(e'),\ldots,\sigma_n(e')]$ be the values of the expression
$e'$ over the input states defined by $f$ (line~\ref{dataset2}).

Given the data $D(x)$ and $D(e')$, \MintHint's goal is to find whether
$x$ and $e'$ are related with each other, that is, whether a change of
value in one is accompanied by a change of value in the other. Because
$D(x)$ contains the \emph{expected (correct)}, rather than current
(erroneous), values of $x$ for the failing tests, the faulty RHS and
its faulty subexpressions should \emph{not} be highly related with
$x$.  Therefore, the expressions that are highly related with $x$ will
be treated as building blocks for the repair. Of course, just because
the fault localization tool marks a statement as potentially faulty,
it does not mean that the statement is actually faulty. The
statistical analysis performed by \MintHint\ is \emph{discriminative}
enough so that, for a spuriously marked statement (\ie~a false
positive), the existing RHS expression $e$ itself may appear as highly
related to $x$.

\paragraph{Statistical Correlation.}

In statistics, the problem of finding whether two variables are
related with each other is called \emph{statistical correlation}.  A
\emph{correlation coefficient} is a numerical measure of the strength
of the correlation between two statistical
variables~\cite{bookCramer,bookRank}.  In our context, a statistical
variable is either the LHS variable $x$ or an expression $e'$.  The
(absolute) value of a correlation coefficient ranges over $[0,1]$. A
correlation coefficient value close to $0$ indicates that the
variables are statistically uncorrelated, whereas values close to $1$
indicate strong correlation.

There are a number of correlation coefficients that are used for
identifying different types of correlations between variables
(\eg~linear or monotonic correlations).  In its current form,
\MintHint\ uses the Spearman coefficient~\cite{bookCramer}. This
coefficient can be applied to any data domain for which there is a
ranking function $r$ that can map the values in the data domain to a
totally-ordered set.  The coefficient is defined even if the data
domains of the two variables are different.  \MintHint\ presently
computes coefficients over three data domains: Booleans, integers, and
ASCII encoding of characters.  In general, there can be multiple
variables in the system. We may want to compute the strength of the
correlation between a pair of variables by eliminating the effect of a
subset $\{c_1,\ldots,c_m\}$ of the other variables. This set is called
a \emph{controlling set}.  The Spearman \emph{partial correlation
  coefficient} is a variant of Spearman coefficient between a pair of
variables with respect to a controlling set.

\vspace{2mm}
\noindent
{\bf Example I.} As an example of how correlation coefficients can be
applied for reasoning about program constructs, consider a dataset
generated by the assignment $y := i + j * k - 10$ over integer
variables $i \in [1,2], j \in [1,\ldots,5], \text{and } k \in [1,\ldots,25].$ 
The Spearman coefficient between the values of the (non-linear)
subexpression $j*k$ and the values of the variable $y$ is
$0.9997$. Clearly, the high coefficient value conforms to the fact
that the term $j*k$ is, numerically, a dominant term on the RHS.
Conversely, $i$ has a very small range and is not a dominant term on
the RHS, so its coefficient is only $0.0230$. Nevertheless, $i$ is
still a contributing term on the RHS. We can use partial correlation
to assess the correlation between $y$ and $i$ by eliminating the
effect of $j*k$. In this example, the value of the partial correlation
between $y$ and $i$, with $j*k$ as the controlling set, is $0.9116$. $\qed$

\vspace{2mm}
\noindent
\emph{Complexity.} The complexity of computing the Spearman coefficient
between datasets of size $n$ is $O(n\;\text{log}\;n)$.  The complexity
of computation of the Spearman partial coefficient is $O(n^3)$, for
$n$ larger than $m$ where $m$ is the size of the controlling set.

\paragraph{Likelihood and Ranking of Expressions.}

The \MintHint\ algorithm is based on the \emph{hypothesis} that an
expression $e'$ is likely to occur on the RHS in the repaired version
of $F$ iff it is highly-correlated with $x$ on the dataset obtained
from the state transformer which gives the expected, correct values
of $x$ even for the failing tests.
Based on this hypothesis, \MintHint\ computes two measures
for expressions: likelihood and partial likelihood.

The \emph{likelihood} of an expression $e'$ to occur in the repaired
RHS, denoted by $\mathsf{likelihood}(e')$, is the absolute value of
the Spearman coefficient between $e'$ and $x$ over the
datasets $D(e')$ and $D(x)$ (obtained at lines~\ref{dataset1} and~\ref{dataset2}).
Given a set $L$ of expressions, the \emph{partial likelihood} of $e'$
to occur in the repaired RHS \emph{along with} the expressions in $L$,
denoted by $\mathsf{p\_likelihood}(e',L)$, is the absolute value of
the Spearman \emph{partial} correlation coefficient of $D(e')$ and
$D(x)$ with $\{D(e'') \mid e'' \in L\}$ as the data of the controlling
set $L$.  

\vspace{2mm}
\noindent
{\bf Example II.} 
Consider a faulty version $y := i + j / k - 10$ of the statement
from Example I, where an incorrect operator $/$ 
is used in place of $*$. Through symbolic execution,
\MintHint\ would compute the expected values of $y$
which would be close, if not identical, to the values
that $y$ would take in the repaired program.
With this data, the expression $j*k$ would get likelihood value of $0.9997$.
On the other hand, the faulty subexpression $j/k$ would get
a likelihood value of $0.2109$. For the sake of this example,
these coefficients are computed wrt values of $y$ from Example I.
\MintHint\ would then use this information
and apply pattern matching to generate a ``{\bf Replace} $j/k$ by $j*k$'' hint.
This technique is discussed next.
$\qed$


\begin{algorithm}[t]
{\small
 \SetAlgoNoLine
 \DontPrintSemicolon
 \SetInd{0.5em}{0.5em}
\KwIn{The faulty statement $F \equiv x := e$, a mapping $D$ from expressions to data, the set $S$ of subexpressions of the faulty RHS,
the set $E$ of enumerated expressions}
\KwOut{A set of simple hints}
 \Begin{
     $R \leftarrow S \cup E$;
   $\SH \leftarrow \emptyset$;    $L \leftarrow \emptyset$\;    
     \label{init}
   \While{ true }{
     \label{while-loop-1}
     \tcp{\small Select the most likely expression}
     $e' \leftarrow {\operatorname{argmax}_{e'' \in R}}\ \mathsf{likelihood}(e'')$; $R \leftarrow R \setminus \{e'\}$\\[1mm]
     \label{select}
     \tcp{{\small Exit loop if likelihood below threshold}}
     \lIf{$\mathsf{likelihood}(e') \leq \delta$}{break}
     \label{loop-exit}
     \tcp{\small Ensure that e' is not subsumed by L}
     \If{$\mathsf{p\_likelihood}(e',L) \geq \beta$}{
       \label{check-p-corr}
       $L \leftarrow L \cup \{e'\}$\;
       \tcp{\small Synthesize a simple hint using e'}
       $(e'', dist) \leftarrow \mathsf{MinEdit}(e', S)$\;
       \label{simple-hint1}
       $\SH \leftarrow \SH \cup \mathsf{GenHint}(e',e'',dist)$\;
       \label{simple-hint2}
     }
   }
   \tcp{\small Remove-hints for unlikely expressions}
   \ForEach{$e' \in S$ for which there is no retain/replace hint}{
     \label{remove-hints}
     \scalebox{0.95}{$\SH \leftarrow \SH \cup \{$($\mathsf{Line}(F)$, {\bf Remove } $e', 1 - \mathsf{likelihood}(e'))\}$}\;
   }
   \Return{$\SH$}
 }
}
 \caption{Algorithm $\mathsf{MintSimpleHints}$}
 \label{algo:mintsimplehints}
\end{algorithm}

\subsection{From Likelihood of Expressions to Hints}
\label{sec:synth}

In this second part, we discuss how \MintHint\ utilizes the likelihood
values of expressions computed in the first part of the algorithm to
synthesize hints. Intuitively, \MintHint\ synthesizes hints by
performing \emph{syntactic pattern matching} between the expressions
that are likely to occur in the repaired RHS and the subexpressions of
the faulty RHS, using different patterns to address different types of
possible faults.
Formally, a \emph{repair hint} $h$ for a statement $F$ is a triple
$(\ell,t,s)$ comprising the line number $\ell$ of statement $F$, a
textual hint $t$, and the hint's score $s$.  The keywords in the
textual hint, shown in bold font, have their
usual English meaning.

\subsubsection{Simple Hints}
\label{sec:simple-hints}

$\mathsf{MintSimpleHints}$ (see Algorithm \ref{algo:mintsimplehints})
synthesizes a set $\SH$ of simple hints given a faulty statement $F$,
a dataset $D$ obtained from $F$'s state transformer, subexpressions
$S$ of the faulty RHS, and extra expressions $E$ over which to search.
The algorithm also makes use of two thresholds $\delta$ and $\beta$,
used respectively, for selecting expressions by likelihood and
for checking partial likelihood of
a candidate expression given the already selected expressions.

$\mathsf{MintSimpleHints}$ initializes the repair space $R$ to $S \cup
E$, the set of simple hints $SH$ and
the set of likely expression $L$ to the empty set (line~\ref{init}). 
It then executes the loop starting at line~\ref{while-loop-1} until
the likelihood of expressions drops below the threshold $\delta$ (line~\ref{loop-exit}).
Within this loop, an expression $e' \in R$ with the highest likelihood
is selected and removed from $R$ (line~\ref{select}).
If the partial likelihood of $e'$ with $L$ as the controlling set
is above the threshold $\beta$ (line~\ref{check-p-corr}) 
then it is added to $L$. This check ensures that $e'$ has sufficient statistical
correlation with $x$ (the LHS) after taking out the effect of $L$,
that is, $e'$ is not subsumed by $L$. In the example in Section~\ref{sec:overview},
once \textsf{s[*i+1] != ENDSTR} is added to $L$, all
expressions until \textsf{s[1] == ENDSTR} (ranked $23$) have
partial correlation below the threshold $\beta = 0.1$.

Now, $\mathsf{MintSimpleHints}$ generates a simple hint for $e'$. 
To do so, it first
invokes function $\mathsf{MinEdit}$, which identifies the expression
$e''$ from $S$ with minimal edit
distance~\cite{DBLP:journals/ipl/ZhangSS92} from $e'$
(line~\ref{simple-hint1}). Let $dist$ be the minimal edit distance.
Then, the algorithm invokes function $\mathsf{GenHint}$, which
synthesizes a simple hint based on the value of $dist$ (line~\ref{simple-hint2}).  If $dist$ is
below a given threshold, it generates hint ``\textbf{Replace} $e''$ by
$e'$''. Conversely, if $dist$ is above such threshold, it generates
hint ``\textbf{Insert} $e'$''.  For the former, the score of the hint
is set to the maximum between the likelihood of $e'$ and the
\emph{un}likelihood of $e''$ (\ie~$1 - \mathsf{likelihood}(e'')$).
For the latter, the score is simply the likelihood of $e'$.  
In the current formulation, \MintHint\
uses threshold of $2$ over the edit distance. This seems to
capture errors resulting from incorrect operator, constant, variable,
array index, and so on. Increasing the threshold may generate
spurious replace hints and reducing it would produce two separate hints,
respectively insert and remove, for $e'$ and $e''$.

If $dist
= 0$ (\ie $e'$ already belongs to $S$), the algorithm generates hint
``\textbf{Retain} $e'$'' and assigns to the hint a score equal to
$\mathsf{likelihood}(e')$. A special case of this is when $e'$ is
exactly the same as the RHS expression for which the algorithm generates
a ``retain the statement'' hint.
Finally, for the expressions that appear
in the faulty RHS but not in set $L$, it generates remove
hints in the loop at line~\ref{remove-hints}, where function
$\mathsf{Line}$ returns the line number of a statement.  The score of
a ``\textbf{Remove} $e'$'' hint is $1 - \mathsf{likelihood}(e')$.


\subsubsection{Compound Hints}
\label{sec:compound-hints}

Function $\mathsf{MintCompoundHints}$, called in Algorithm~\ref{algo:minthint} at
line~\ref{compound-hints}, computes compound hints using the same
inputs as $\mathsf{MintSimpleHints}$.  Due to space constraints, we do
not typeset $\mathsf{MintCompoundHints}$ and simply explain the key
similarities and differences between $\mathsf{MintCompoundHints}$
and Algorithm~\ref{algo:mintsimplehints}.
Similar to Algorithm~\ref{algo:mintsimplehints}, it iteratively computes
the set of likely expressions. However, it uses $\mathsf{p\_likelihood}(e'',L)$
in place of $\mathsf{likelihood}(e'')$ at line~\ref{select}
and the branch predicate at line~\ref{check-p-corr} is replaced with $true$.
The value of partial likelihood gives the measure of the likelihood
of $e''$ to appear in the repaired RHS \emph{along with} 
the expressions in $L$. 
As a consequence, the expressions that get added to the set $L$
are all those expressions which can occur \emph{together} in the repaired RHS.
In contrast, the expression selected at line~\ref{select} in
Algorithm~\ref{algo:mintsimplehints} is selected only based on its
\emph{individual} likelihood.

In general, $\mathsf{MintCompoundHints}$ can generate more than one
set of likely expressions.
More specifically, if multiple expressions have the highest
partial correlation coefficient (at line~\ref{select})
in the first iteration of the loop, 
then the algorithm would partition them into three sets based on their edit
distance from the faulty RHS and its subexpressions: (1)~equal to
zero, (2)~less than equal to $2$, and (3)~more than $2$. 
It initializes three sets of likely expressions by selecting
one expression from each of the partitions above (if not empty).
It then proceeds independently to select other expressions
to add to each of them.
(Conversely, when generating simple hints, for multiple expressions with
the highest correlation coefficient, $\mathsf{MintSimpleHints}$ selects one
expression at random without partitioning them by edit distance.)
For example, in Section~\ref{sec:overview}, the expression
\textsf{s[*i+1] == ENDSTR} is the only expression, among the seven
expressions with the highest likelihood, which is at edit distance 
less than equal to $2$. 

For each expression in the set $L$, $\textsf{MintCompoundHints}$
generates a hint with the pattern matching logic used at line~\ref{simple-hint2}
in Algorithm~\ref{algo:mintsimplehints}. We call each of these hints,
a \emph{constituent hint}, of the compound hint.
We say that a pair of constituent hints of a compound hint
\emph{conflict} if they refer to either the same or overlapping
subexpressions of the faulty RHS.  Two subexpressions of the faulty
RHS overlap if their subtrees in the AST of the faulty RHS have some
common node(s).  Overlapping subexpressions can be identified by a
simple walk over the AST.  In order to ensure that no conflicting
hints are generated, \MintHint\ removes a subexpression and all
expressions that overlap with it from the repair space as soon as the
subexpression is added to set $L$.  (Simple hints are independent of
each other and hence are permitted to conflict.)  
Finally, analogous to the loop at line~\ref{remove-hints} in Algorithm~\ref{algo:mintsimplehints},
$\mathsf{MintCompoundHints}$ generates remove hints for the
subexpressions of the faulty RHS for whom there is no retain
or replace hint synthesized earlier.

There could be multiple simple hints that suggest
retention of different subexpressions of the RHS. In these cases,
these hints are clustered into one single compound hint.
The score of a compound hint is the maximum of the scores of
the constituent hints. 


%% file: eval.tex
\section{Evaluation}
\label{sec:eval}

To assess the effectiveness of our approach, we implemented 
\MintHint\ to synthesize repair hints for C programs
and performed an empirical evaluation. Our implementation uses 
Zoltar~\cite{zoltar} for fault localization, 
KLEE~\cite{DBLP:conf/osdi/CadarDE08} for dynamic symbolic execution
with constraint solving,
and Matlab\footnote{\url{http://www.mathworks.com/products/matlab/}} 
for statistical analysis.

Our evaluation consists of two parts. We performed a \emph{user study}
to evaluate whether \MintHint\ improves developers' productivity.
The user study is augmented with the application of hints 
on a \emph{case study}.
We investigate the following research questions:

\textbf{RQ1:} \emph{Usefulness of hints} --- Can \MintHint\ produce useful
hints, enabling developers to repair programs more effectively? 

\textbf{RQ2:} \emph{Robustness} --- How does \MintHint\ perform when
the repair space is incomplete (\ie does not contain the 
repaired version of the faulty expression)
and is supplied imprecise data?

\textbf{RQ3:} \emph{Performance and scalability} --- How does \MintHint\
scale to large programs, state transformers, and repair spaces?

\subsection{User Study}
\label{sec:user-study}

\begin{table}
\centering
{\small
\begin{tabular}{|l|c|c|}
\hline
\emph{Program} & \emph{LOC} & \emph{\#Tasks} \\
\hline
\textsf{print\_tokens2} & $570$ & $2$ \\
\hline
\textsf{replace} & $564$ & $4$ \\
\hline
\textsf{tcas} & $173$ & $4$ \\
\hline
\end{tabular}
}
\vspace{-8pt}
\caption{Description of tasks.}
\vspace{-8pt}
\label{table:tasks}
\end{table}

\begin{table*}
{\small
\begin{tabular}{clllccc}
\toprule
\emph{Task} & \emph{Program-Version} & \emph{Nature of fault}     & 
\emph{Type of the most useful hint} & \emph{Rank of the most} & \emph{\#Total hints} & \emph{\#Stmts with only}\\
&     & 
&  & \emph{useful hint}  &               & \emph{``Retain the stmt'' hints} \\
\midrule
$1$ & \textsf{print\_tokens2-v6} & Incorrect array index 
& Remove & $10$ & $31$ & -- \\
$2$ & \textsf{replace-v7} & Superfluous expression 
& Compound (Remove + others) & $4$ & $13$ & $1$ \\
$3$ & \textsf{print\_tokens2-v7} & Superfluous expression 
& Compound (Remove + others) & $4$ & $8$ & -- \\
$4$ & \textsf{replace-v18} & Missing expression 
& Compound (Insert + others) & $1$ & $3$ & $1$ \\
$5$ & \underline{\textsf{replace-v23}} & Incorrect array index 
& Replace & $1$ & $6$ & $3$ \\
$6$ & \textsf{tcas-v28} & Incorrect operator 
& Replace & $6$ & $26$ & -- \\
$7$ & \textsf{replace-v8} & Missing expression 
& Compound (Retain + others) & $1$ & $6$ & $2$ \\
$8$ & \underline{\textsf{tcas-v2}} & Incorrect constant 
& Replace & $7$ & $13$ & $1$ \\
$9$ & \underline{\textsf{tcas-v1}} & Incorrect operator 
& Replace & $2$ & $11$ & $3$ \\
$10$ & \underline{\textsf{tcas-v12}} & Incorrect operator 
& Compound (Retain + others) & $11$ & $14$ & -- \\
\bottomrule
\end{tabular}
}
\vspace{-8pt}
\caption{Description of faults in the user study and of useful hints identified by the users:
The tasks that could not be finished in the control phase (without hints) are underlined.
All the tasks were completed in the experimental phase (with hints).}
\vspace{-8pt}
\label{table:hints}
\end{table*}

\paragraph{Experimental Setup.}
We performed fault localization on the programs from the 
Siemens suite~\cite{siemens}. The Siemens suite consists
of a few programs with multiple faulty versions for each of them.
Table~\ref{table:tasks} lists the programs and the number of faulty versions
that were selected as tasks. In all, $10$ tasks were selected.
In the user study, each user was required to work on two 
independent tasks. To keep each task manageable within $2$h,
for each task, only the top $5$ statements
identified by Zoltar as potentially faulty were presented to the users. For each of the chosen
tasks, the actual faulty statement belonged to this list. 
The tasks represent a diverse collection of faults (see Table~\ref{table:hints}).
%
For each program and candidate faulty statement,
the state transformers were obtained for failing tests through
symbolic execution with the timeout of $5$m per test. 
In one of the tasks, 
symbolic execution of many failing tests timed out. Consequently,
the data from the passing tests was seen to bias the results.
Therefore, for that task, only half of the passing tests were used.

Of the $10$ users who participated in our study, $8$ were working professionals
and $2$ were graduate students (with prior industry experience) 
-- none affiliated with our research group.
Of these, $8$ stated that they had moderate to high expertise with C programming/debugging
with everyone having at least $1$ year of experience in C programming.
For each task, the input/output specifications
of methods, and meaning of variables and named constants
were presented as comments in the source code. In addition, each user was provided
a test suite of $10$ passing tests and 
was given $15$m to study the program before starting with program repair.

\begin{table}
\centering
{\small
\begin{tabular}{@{}|@{\;}p{3.3cm}@{\;}|@{\;}p{2cm}@{\;}|@{\;}p{2.3cm}@{\;}|@{}}
\hline
& \multicolumn{1}{c@{\;}|@{\;}}{\emph{Control phase}} & \multicolumn{1}{c@{\;}|}{\emph{Experimental phase}} \\
& \multicolumn{1}{c@{\;}|@{\;}}{(without hints)} & \multicolumn{1}{c@{\;}|}{(with hints)}\\
\hline
\multicolumn{3}{|c|}{\cellcolor[gray]{0.8}{\textbf{Quantitative Analysis}}} \\
\hline
\emph{Successful localization} & $8/10$ & $10/10$\hspace{7mm} $(\mathit{+2})$ \\
\emph{Successful repair} & $6/10$ & $10/10$\hspace{7mm} $(\mathit{+4})$ \\
\emph{Avg. time to repair} & $91$m + $4$ timeouts & $29$m (no timeouts)\\
\emph{Avg. speedup (excl. timeouts)} & NA & $5.8$x\\
\hline
\multicolumn{3}{|c|}{\cellcolor[gray]{0.8}{\textbf{Qualitative Analysis (Ratings given by the users)}}} 
\\
\hline
\multirow{3}{*}{\emph{Difficulty of localization}} & Easy: $2$ & Easy: $6$ \hspace{5mm} $(\mathit{+4})$ \\
& Moderate: $6$ & Moderate: $3$ \\
& Difficult: $2$ & Difficult: $1$ \\
\hline
\multirow{3}{*}{\emph{Difficulty of repair}} & Easy: $3$ & Easy: $7$ \hspace{5mm} $(\mathit{+4})$ \\
& Moderate: $5$ & Moderate: $1$ \\
& Difficult: $2$ & Difficult: $2$ \\
\hline
\end{tabular}
}
\vspace{-8pt}
\caption{Results of the user study.}
\vspace{-8pt}
\label{table:results}
\end{table}

There were two phases of the user study.
In the \emph{control phase}, the users were given the fault localization information and test suite.
In the \emph{experimental phase}, in addition, they were given the repair hints.
Each user worked on a single task in a phase and was given $2$h to complete the task. 
A task was considered to be \emph{complete} if the repaired program passed all the tests.
The users chose the programs for the control phase by drawing lots. 
We had mapped each task in the control phase to a task in the experimental phase with
the objective of avoiding a user working on the same task or another
version of the same program in both phases. 
A presentation was made to them to explain
the meaning of different types of repair hints.
The users were provided Linux based machines
with a debugger, text editors and IDEs, and the standard command line utilities.

\paragraph{(RQ1) Usefulness of hints.}

Table~\ref{table:results} summarizes the results of the user study.
In the control phase (without hints),
out of $10$ tasks, the users could localize the faults
to the actual faulty statement in $8$ cases but only managed to repair the
programs in $6$ cases. On the contrary, in the experimental
phase (with hints), the users could perform localization and repair in all the $10$ cases.
Further, the average time taken to repair a fault was
$91$m in the control phase (excl. the $4$ timeouts), 
whereas it was $29$m in the experimental phase. The average speedup obtained
with the use of hints, for the $6$ tasks that were completed in both the phases,
was equal to $5.8$x.
The users were asked to rate the difficulty
level of fault localization and repair for their tasks 
as easy, moderate, or difficult. 
The ratings, \emph{for the same set of tasks}, differ across the two phases.
Notably, with the hints, $4$ more tasks were rated by the users as easy.
These \emph{qualitative ratings} corroborate the quantitative results
presented above.

\begin{table}
{\small
\begin{tabular}{|l|l|c|}
\hline
& \emph{Yes} & \emph{No} \\
\hline
\emph{Incompleteness of repair space} & $5$ & $5$ \\
\hline
\emph{Noise in state transformers obtained from tests} & $7$ (max. $27$\%) & $3$ \\
\hline
\end{tabular}
}
\vspace{-8pt}
\caption{Count of tasks in the user study wrt incompleteness of the
repair space and noise in state transformers.}
\vspace{-8pt}
\label{table:noise}
\end{table}

The users uniformly mentioned that the hints were useful
and were asked to indicate the hint that was most useful to them.
The ranks of the most useful hints in Table~\ref{table:hints} are over
the entire list of hints for all statements presented to the user
for that task. Note that \MintHint\ may produce multiple hints
for a faulty statement.
The most useful hints were all in top $10$ except for one case.
The last column in Table~\ref{table:hints} gives the number of
statements for which \emph{only} a ``retain the statement'' hint was generated.
These are the statements that are classified by \MintHint\ as
unlikely to be faulty. Across the $10$ tasks and $50$ statements
in the fault localization lists, $40$ statements are likely to be false positives
($10$ are true positive, \ie definitely faulty).
Out of these, $11$ statements (more than $25$\%) are eliminated by \MintHint.
This contributes greatly to ease of localization.
We believe that there is no obvious way to rectify these statements.
In fact, no user in the study (even in the control phase) came up with a repair 
for any of these statements.

For $6$ tasks (tasks 2--6 and 9), the most useful hint
had the \emph{precise} information required to repair the fault.
For task 10, the faulty RHS was of the form \textsf{exp1 || exp2} and the compound
hint suggested that both \textsf{exp1} and \textsf{exp2} be retained
but did not say the same for the entire RHS. The user therefore
suspected that the operator was incorrect and correctly replaced
it with \textsf{\&\&}. For task 1, the hint suggested
only removal of the incorrect expression. The user had to come up 
with the substitute expression. In task 7, the hint did not
suggest the expression to be inserted. The user mentioned that
the hints helped mainly in localizing the fault. This is possible
because \MintHint\ eliminates $2$ other statements in this case
(see the last column for task 7 in Table~\ref{table:hints}). 
In the case of task 8,
the replace hint did not suggest the required (named) constant to be used.
Nevertheless, the user observed that after substituting the new
expression suggested in the hint, many failing tests started passing
and subsequently inferred the right constant manually. 
Of these, tasks 5, 8, 9, and 10 were \emph{not} completed in the control phase.

The number of total hints per task given in Table~\ref{table:hints}
depends on the following factors:
(1)~The thresholds on correlation coefficients
which determine how many expressions
end up in the set of likely expressions, thereby, also affect the
number of hints. Across all tasks, the thresholds on correlation coefficient
and partial correlation coefficient for generation of simple hints were $0.4$
and $0.1$ respectively. For compound hints, it was $0.6$.
(2)~If the symbolic execution times out on all failing tests
for a statement then the hint generation algorithm is
not run for that statement (see Section~\ref{sec:state-trans}).

Though the study is not large
enough to measure the difference with statistical significance,
the results suggest that repair hints can contribute 
significantly in improving developers' productivity and 
even the feasibility of repair (within limited time).

\paragraph{(RQ2) Robustness.}

In practice, it is difficult to estimate the syntactic space to
search for a complete repair. In our experiments, for each statement, 
apart from the subexpressions of the potentially faulty expression, 
expressions of size up to $4$ (over the variables in scope
at the statement) were added to the repair space.
The size of an expression is the number of nodes in its
abstract syntax tree (AST). For expressions involving arrays,
an occurrence of an array with the index expression is counted as size $1$
and the index expressions themselves can go up to size $4$.
We call a repair space \emph{complete}
only if the repaired version of the expression belongs to it.
For example, in task 10, the repaired expression
\textsf{exp1 \&\& exp2} was not in the repair space. 

\MintHint\ derives state transformers for failing tests by symbolic execution.
In some cases, the derived constraints may have
multiple satisfying assignments but not all of them can be
observed in the execution of the repaired version of
the program. The constraint solver may pick any one of them.
For passing tests, the state transformer is obtained
by concrete execution. Even though the test passes, the value
generated by the faulty expression may not be observed in the repaired
version. These situations make the resulting data, which is used as a specification,
imperfect (noisy) and may in general invalidate the applicability of a repair.

\begin{table}
{\small
\begin{tabular}{c|c|c|c|c|}
\multicolumn{2}{c}{} & \multicolumn{3}{c}{\emph{\#Exprs in repair space}} \\
\multicolumn{2}{c}{} & \multicolumn{3}{c}{(\#Columns of the data matrix)} \\
\hhline{~----}
\multirow{2}{*}{\begin{minipage}{1.5cm}{\emph{Size of state transformer}}\end{minipage}}
& & \emph{Upto $5$}k & \emph{Upto $10$}k & \emph{> $10$}k\\
\hhline{~----}
& \emph{Upto $1$}k & {\cellcolor[gray]{0.8}} < $1$m & {\cellcolor[gray]{0.8}} < $1$m & < $1$m \\
\hhline{~----}
\multirow{2}{*}{\begin{minipage}{1.6cm}{(\#Rows of the data matrix)}\end{minipage}}
& \emph{Upto $10$}k & {\cellcolor[gray]{0.8}} < $1$m  & {\cellcolor[gray]{0.8}} < $5$m & < $5$m \\
\hhline{~----}
& \emph{ > $10$}k & < $1$m & {\cellcolor[gray]{0.8}} < $5$m & $\approx 1$h \\
\hhline{~----}
\end{tabular}
}
\vspace{-8pt}
\caption{Performance and scalability chart: Tasks
from the user study belong to all cells except cell at position $(3,2)$.
Tasks from the case study
(Section~\ref{sec:case-study}) belong to the shaded cells.
}
\vspace{-8pt}
\label{table:scalability}
\end{table}

Table~\ref{table:noise} gives the count of tasks which had noise in
the state transformers and where the repair space was incomplete.
This information is provided only for the actual faulty statement. 
To estimate the amount of noise in the data,
we first obtain the \emph{noise-free} state transformer of the (known) repaired version
independently by concrete execution over the test suite. 
An entry in the state transformer
obtained over the faulty version is classified as \emph{noisy}
if it does not belong to the noise-free state transformer.
There were $7$ tasks with noisy data with
the maximum of $27$\% noise in one of them and there were $5$ tasks where the
actual repaired expression did not belong to the repair space. Nevertheless,
the successful completion of the tasks in the user study 
indicates that useful hints could be synthesized even in these challenging cases.
The key reasons for this are (1)~the use of statistical correlation
which is robust in presence of noise and (2)~the ability of \MintHint\
to synthesize compound hints from building blocks. In particular,
the repair spaces were incomplete for tasks 2--4, 7, and 10,
and as Table~\ref{table:hints} shows, in each of these cases,
the most useful hint was a compound hint.

\paragraph{(RQ3) Performance and scalability.}

The complexity of statistical correlation computation dominates
the cost of hint generation (see Section~\ref{sec:ranking} for 
discussion on its complexity).
It works on a two-dimentional matrix where the number of rows is equal to 
the size of the state transformer (the number of input/output pairs)
and the number of columns is equal to the number of
expressions in the repair space.
Table~\ref{table:scalability} gives the performance and
scalability chart summarizing all runs of the hint generation algorithm
for all tasks and faulty statements in the user study.
Despite the large datasets, the hint generation algorithm scales well. 
Zoltar took slightly over $2$m on an average for fault localization.
Except for a few statements, KLEE
ran to completion on all the failing tests 
within $5$m (the timeout set by us). The timings are measured on a 
desktop with Intel i5 CPU@3.20 GHz and 4GB RAM.

\subsection{A Case Study}
\label{sec:case-study}

\begin{table}[t]
\centering
{\small
\begin{tabular}{@{}c@{\;}l@{\;}l@{\;\;}l@{\;\;}c@{\;}c@{\;}c@{}}
\toprule
\emph{Task}  & \emph{Version-} &     &
\emph{Type of} & \emph{Rank of} & \emph{\#Total} & \emph{\#Stmts with}\\
   & \emph{Fault} & \emph{Nature of fault} 
& \emph{useful} & \emph{useful}  &             \emph{hints}  & \emph{``Retain stmt''}\\
& & & \emph{hint} & \emph{hint} & & \emph{hint only}\\
\midrule
$1$ & \textsf{v2-f1} & Incorrect operator 
& Replace & $1$ & $11$ & $10$ \\
$2$ & \textsf{v3-f4} & Incorrect constant 
& -- & -- & $11$ & $3$ \\
$3$ & \textsf{v3-f6} & Incorrect constant 
& Replace & $1$ & $10$ & $9$ \\
$4$ & \textsf{v5-f1} & Incorrect operator 
& Replace & $13$ & $13$ & $12$ \\
$5$ & \textsf{v6-f1} & Incorrect \emph{operators} 
& Compound & $2$ & $4$ & -- \\
$6$ & \textsf{v6-f2} & \multicolumn{5}{c}{Symbolic execution times out with $15$m threshold}\\
\bottomrule
\end{tabular}
}
\vspace{-8pt}
\caption{Description of faults and hints in the case study.}
\vspace{-8pt}
\label{table:case-study-tasks}
\end{table}

\begin{table}
{\small
\begin{tabular}{|l|l|c|}
\hline
& \emph{Yes} & \emph{No} \\
\hline
\emph{Incompleteness of repair space} & $1$ & $4$ \\
\hline
\emph{Noise in state transformers obtained from tests} & $4$ (max. $97$\%) & $1$ \\
\hline
\end{tabular}
}
\vspace{-8pt}
\caption{Count of tasks in the case study wrt incompleteness of the repair space
and noise in state transformers.}
\vspace{-8pt}
\label{table:case-study-noise}
\end{table}

\paragraph{Experimental Setup.}
We applied \MintHint\ on a commonly used Unix utility program,
the stream editor \textsf{sed} obtained from the SIR repository~\cite{SIR}. 
\textsf{sed} is a reasonably
large program with over $14$K LOC and $250$ functions.
The SIR repository consists of several versions of \textsf{sed} seeded
with different faults. We performed fault localization on them using Zoltar.
Table~\ref{table:case-study-tasks} lists the versions and fault-IDs of the programs
that were selected as tasks. In all, $6$ tasks were selected.
The top $15$ statements identified by Zoltar as potentially faulty were
considered for repair. For each of the chosen
tasks, the actual faulty statement belonged to this list. 
For each version and candidate faulty statement,
the state transformers were obtained for failing tests through
symbolic execution with the timeout of $5$m per test for the first
three tasks and $15$m for the remaining. 
In the sixth task, symbolic execution timed out for all failing tests
and hence it was not subjected to further analysis.

\paragraph{(RQ1) Usefulness of hints.}

For each of the tasks, Table~\ref{table:case-study-tasks} shows the
nature of the fault and the type of the most useful hint. 
For tasks 1, 3, and 4, a replace hint synthesized by \MintHint\ when
applied, immediately lead to success on both passing and failing tests.
Task 5 consists of two faults in the same statement. A compound
hint suggested removal of two subexpressions
and retention of the non-faulty subexpression. 
We removed the subexpressions identified by \MintHint\ for removal and 
obtained well-formed expressions by removing
operators around them. 
This change however
lead to only a \emph{partial repair} since \MintHint\ did not
generate the expressions that should be used for replacing the
faulty ones. 
With this change, all the previously passing tests
continued to pass and several previously failing tests too
started to pass. In task 2, \MintHint\ did not produce any useful
hint for the faulty statement (due to excessive noise, as discussed below).

Interestingly, \MintHint\ generated only ``retain the statement'' hints
for many statements identified by Zoltar. 
Across the $5$ tasks and $75$ statements
in the fault localization lists, $70$ statements are likely to be false positives
($5$ are true positive, \ie definitely faulty).
Out of these, $34$ statements (more than $45$\%) are eliminated by \MintHint.
In fact, in tasks 1, 3, and 4, retain were the only other type
of hints apart from the replace hint required for program repair.
We studied the faulty statements which had only the retain hints but could
not identify a way to change them to repair the faults.

This study gives preliminary evidence that \MintHint\ can be
applied to large programs with not so accurate fault localization lists.

\paragraph{(RQ2) Robustness.}
Repair spaces were constructed in a manner similar to the
user study and with the same bound on expression sizes.
As shown in Table~\ref{table:case-study-noise}, the repair space
searched by \MintHint\ contained the repaired version of the faulty RHS
in all but one case. The data obtained for the failing tests however
contained noise in all cases except one. Our approach for estimating
the amount of noise is explained in Section~\ref{sec:user-study}.
In one case (task 2), the noise was $97$\% and \MintHint\
could not produce any useful hint.

\paragraph{(RQ3) Performance and scalability.}
In Table~\ref{table:scalability}, the shaded cells denote
the time taken by \MintHint\ for statistical correlation analysis
and hint generation for the tasks in the case study.
No task took more than $5$m for these steps. Fault localization
finished within $1$m on an average for these tasks. 
While symbolic execution completed within $5$m 
on most failing tests for the first three tasks
it required $15$m for most cases for tasks 4 and 5.
However, even with $15$m threshold, 
it timed out for all failing tests and all potentially
faulty statements for the sixth task. 

\subsection{Limitations and Threats to Validity}
\label{sec:threats}

One of the main limitations of our approach is its reliance on
symbolic execution for deriving state transformers, which is a complex
and expensive technique. 
However,
these techniques are becoming increasingly efficient, and
many of their practical limitations are being addressed
(\eg
\cite{sinha2012alternate, godefroid2010compositional}). 
Moreover, as we discuss in
Section~\ref{sec:conclusions}, we plan to investigate 
alternative, less expensive ways to build state transformers.
Further, it is technically difficult to obtain only those
values which can be observed in the repaired program through symbolic execution.
This makes the state transformers noisy. Due to the statistical reasoning
applied in \MintHint, it produced useful hints even in presence of 
noise in many cases. The so-called measurement
errors leading to noisy data are common in other application domains 
as well and a large body of work, commonly
called \emph{outlier detection}, exists to deal with them 
(see~\cite{chandola2007outlier} for a survey). We plan to investigate
applications of these techniques to further improve tolerance of \MintHint\ to noise.

Like every empirical evaluation, ours too has potential threats to validity.  
Threats to \emph{internal validity} for the user study include 
\emph{selection bias} where the users working on the same task
in control and experimental phases may have different expertise
and \emph{testing bias} where activities before the study may affect 
the outcome. Only two users had indicated low expertise with
C programming, to prevent selection bias, we paired them in
such a way that their tasks were interchanged in the two phases.
Other users picked their tasks randomly.
To mitigate testing bias, we ensured that no user worked on
the same task or two faults of the same program, in the two phases
of the user study. Further, we only made a presentation about
the meaning of repair hints to them and did not provide any
hands-on tutorial.  There may be faults in our implementation that might have
affected our results. To address this threat, we manually checked many
of our results and did not encounter any error. 

Threats to \emph{external validity} arise because our results may
not generalize to other group of developers (in the case of user study)
and program repair tasks. In the user study, we ensured that
the users did not have any prior experience with the programs
used as repair tasks. While this ensures a level playing field,
it leaves out users who might have better familiarity with the
programs. The tasks in the user study were \emph{not} hand-picked.
We applied a well-defined criterion for their selection. The programs in Siemens
suite were sorted by the rank of the actual faulty statement
in the respective localization lists. Within each rank, the program-fault names
were then sorted in the lexicographic order and finally, the first two tasks at each 
fault localization rank were selected. Similarly, the tasks in the case study,
involving \textsf{sed}, consist of all tasks from the SIR repository~\cite{SIR}
with only a single
faulty statement such that the statement occurs within top $15$
statements returned by a third-party fault localization tool, Zoltar.
A few tasks could not be included because of limitations of the symbolic execution tool.
The performance of \MintHint\ is a function of the test suite also.
It will take a much larger evaluation to ascertain how the quality of
tests affects the quality of repair hints.
It is however important to note that the tasks and test suites we used were also used in numerous
previous papers in the area (\eg \cite{bugassist, nguyen2013, Zhang2006}). 
Nevertheless, the number of
tasks considered is small, so our findings may
not generalize to other programs or faults. We plan to perform an
additional extensive empirical evaluation to confirm our current results.



%% file: relatedwork.tex
\section{Related work}
\label{sec:related-work}


Early work by Arcuri~\cite{arcuri2008}, later extended by Le Goues and
colleagues~\cite{legoues-tse2012}, proposes the use of genetic
programming to automatically generate repairs that make failing test
cases pass and do not break any passing test case.  Debroy and Wong
propose a similar approach, but based on the use of mutation~\cite{debroy2010using}. 

Other approaches rely on the use of program specifications to repair
the code. Pei and colleagues propose an approach for finding program
repairs given program contracts~\cite{pei2011code, wei10}.  Jobstmann,
Griesmayer, and Bloem use a game theoretical approach for identifying
repairs that satisfy a linear-temporal-logic specification of the
program to be repaired~\cite{jobstmann2005program}.  Gopinath, Malik,
and Khurshid's approach builds a SAT formula that encodes the
constraints imposed by the specification on the program behavior and,
if the formula is satisfiable, derives a repair from the SAT
solution~\cite{DBLP:conf/tacas/GopinathMK11}.  Konighofer and Bloem
present a template-based approach that, given a faulty program and a
specification, performs a symbolic analysis of program inputs and
template parameters to generate a
repair~\cite{konighofer2011automated}. He and Gupta propose a
technique that computes the weakest preconditions along a failing
trace and compares the computed conditions with functions' pre- and
post-conditions to find and correct faults~\cite{DBLP:conf/fase/HeG04}.
Logozzo and Ball's approach generates verified program repairs from
failed verification checks of programs that have developer-supplied
modular specifications~\cite{Logozzo:2012:MVA:2384616.2384626}.

Yet other approaches leverage the existing test suite to infer 
specifications and generate repairs
accordingly.  \textsc{pachika}~\cite{dallmeier2009generating} models a
program's behavior for passing and failing test cases and generates a
repair based on the differences between these models of correct and
incorrect behavior. In a recent paper, Nguyen and colleagues
propose the SemFix approach that combines angelic debugging~\cite{chandra2011}
and program synthesis~\cite{jha2010oracle} to automatically identify
program repairs~\cite{nguyen2013}.  
BugFix~\cite{jeffrey2009bugfix} shares with
our approach the idea of providing developers with ``bug-fix
suggestions'', rather than actual repairs. Unlike our approach, however,
BugFix generates suggestions using a machine-learning
approach based on knowledge acquired from previous bug repairs.

Finally, some approaches perform program repair in specific domains,
such as repairs for data structures~\cite{demsky2006inference}, web
application repairs~\cite{samimi2012automated}, and repairs targeted
at security vulnerabilities~\cite{perkins2009automatically} or
concurrency faults~\cite{Jin2012concurrency}.



%% file: conclusion.tex
\section{Conclusions and Future Work}
\label{sec:conclusions}

We presented \minthint, a novel technique for semi-automated program
repair. 
The key novelty of our approach is that it combines
symbolic, statistical, and syntactic reasoning to synthesize repair hints,
given only the faulty program and a test suite.  
Our evaluation of \minthint\ provides initial but strong evidence
that our approach is effective.
%
An interesting direction for future work is the investigation of how repair
hints could be used to further automate the program-repair process.
For example, we envision that hints could be used to inform program
synthesis (\eg \cite{gulwani2011synthesizing, gulwani2011automating,
  jha2010oracle}) or sketching (\eg
\cite{DBLP:conf/asplos/Solar-LezamaTBSS06, solar2005programming}).
In the future, we will extend our technique so that it can handle the
more challenging case of faults involving multiple statements. 
Finally, in order to improve performance of \MintHint,
we plan to investigate (1)~alternative, more efficient techniques for
building operational specifications and (2)~outlier detection
mechanisms to further improve tolerance of \MintHint\ to noise.
